\documentclass[twocolumn,amsmath,amssymb,amsfonts,superscriptaddress,floatfix,showpacs,prd,aps]{revtex4}

\newcommand{\del}{\partial}
\newcommand{\Tr}{\mathrm{Tr}}

\newcommand{\vev}[1]{\langle{#1}\rangle}

\usepackage{graphicx,mathrsfs,flafter}
\usepackage{color}
\usepackage{hyperref}

\begin{document}
\title{
Vacuum fluctuations and the thermodynamics of  chiral  models }
\author{V. Skokov}
\affiliation{%
GSI Helmholtzzentrum f\"ur Schwerionenforschung, D-64291
Darmstadt, Germany}
\author{B.~Friman}
\affiliation{%
GSI Helmholtzzentrum f\"ur Schwerionenforschung, D-64291
Darmstadt, Germany}
\author{E.~Nakano}
\affiliation{%
GSI Helmholtzzentrum f\"ur Schwerionenforschung, D-64291
Darmstadt, Germany}
\affiliation{%
Physics Division, Faculty of Science, Kochi University, Kochi 780-8520, Japan}
\author{K.~Redlich}
\affiliation{%
Institute of Theoretical Physics, University of Wroclaw, PL--50204 Wroc\l aw, Poland}
\affiliation{%
Theory Division, CERN, CH-1211 Geneva 23, Switzerland}
\author{B.-J.~Schaefer}
\affiliation{%
Institut  f\"ur Physik, Karl-Franzens-Universit\"at, A-8010 Graz, Austria
}


\date{\today}

\begin{abstract} 
  We consider the thermodynamics of chiral models in the mean-field
  approximation and discuss the relevance of the (frequently omitted)
  fermion vacuum loop. Within the chiral quark-meson model and its
  Polyakov loop extended version, we show that the fermion vacuum
  fluctuations can change the order of the phase transition in the
  chiral limit and strongly influence physical observables. We compute
  the temperature-dependent effective potential and baryon number
  susceptibilities in these models, with and without the vacuum term,
  and explore the cutoff and the pion mass dependence of the
  susceptibilities. Finally, in the renormalized model the
  divergent vacuum contribution is removed using the dimensional
  regularization.
\end{abstract}

\maketitle

\section{Introduction}

The critical properties of strongly interacting matter near the chiral
transition at finite temperature and density are studied in first
principle calculations using lattice simulations of QCD~\cite{lgt}. A
complementary approach is offered by effective chiral models
\cite{Stephanov:2007fk, njl, cs, Scavenius:2000qd, Schaefer:2009ui,
  fukushima, PNJL, Schaefer:2007pw, Schaefer:2008hk, qm}, like the
Nambu-Jona-Lasinio (NJL) \cite{njl} and the chiral quark-meson
(QM)~\cite{qm} models as well as the corresponding extended versions,
which include the interaction of quarks with a uniform temporal gluon
field. The latter, the Polyakov loop extended Nambu--Jona--Lasinio
(PNJL)~\cite {fukushima, PNJL} and quark-meson
(PQM)~\cite{Schaefer:2007pw, Schaefer:2009ui} models reproduce
prominent features of QCD thermodynamics, obtained in lattice
simulations. In particular, both models exhibit a chiral as well as a
deconfinement transition.

In this paper, the consistency of mean-field calculations of many-body
systems within chiral effective models like the NJL/PNJL and QM/PQM
models is addressed. In particular, we discuss the role of the lowest
order fermion vacuum contribution. In the NJL/PNJL model, this term is
responsible for the dynamical breaking of the chiral symmetry in
vacuum and is therefore necessarily taken into account. On the other
hand, in the QM/PQM model, the spontaneous breaking of the chiral
symmetry is due to the meson potential. Hence, for qualitative
considerations, the fermion vacuum loop is frequently omitted~\cite
{Scavenius:2000qd, Schaefer:2006ds, Bowman:2008kc, Gupta:2009fg,
  Nickel:2009wj, Kahara:2009sq, Kapusta:2009yd}.

We show that a mean-field approximation, where the fermion vacuum term
is neglected, commonly referred to as the no-sea approximation, leads
to a distortion of the critical behavior at the chiral transition. In
particular, in this approximation the order of the transition in the
chiral limit is {\em always} first order. In contrast, when this term
is included, the transition is of first or second order, depending on
the choice of coupling constants and on the baryon density.

Differences in the thermodynamics between the NJL and QM models found
in the literature can be attributed to the use of the no-sea
approximation in the QM model. For instance, in this approximation,
the adiabatic trajectories obtained in the QM model exhibit a kink at
the chiral crossover transition, while in the NJL model they are
smooth everywhere~\cite{Scavenius:2000qd}. This effect is a trace of
the underlying first-order phase transition of the QM model in the
chiral limit, when fermionic vacuum fluctuations are
neglected~\cite{Nakano:2009ps}. Moreover, the omission of fermion
vacuum fluctuations can result in a violation of the semi-positivity
of the spectral function as well as in a violation of detailed
balance~\cite{fujii}. The vacuum fermion fluctuations  were also shown in
the PQM model to considerably affect the structure of the phase diagram in
the presence of an external
 magnetic field~\cite{Mizher:2010zb}.

In this paper we demonstrate the importance of fermion vacuum
fluctuations for a consistent formulation of the thermodynamics of the
QM and PQM models in the mean-field approximation. We show that, in
order to reproduce the second-order chiral phase transition expected
for two flavor QCD within these models~\cite{Pisarski:1983ms}, it is
necessary to take the fermion vacuum term in the thermodynamic
potential into account and to perform a proper regularization of the
divergence. Furthermore, we also show that when the chiral symmetry is
explicitly broken, the thermodynamics is strongly affected by the
no-sea approximation. Finally, we remove the infinities of the vacuum
term using  the dimensional regularization.

\section{Thermodynamics of the quark-meson model}

The chiral QM model is used as an effective realization of the
low--energy sector of QCD. The replacement of the $SU(N_c)$ local
gauge invariance of QCD by a global symmetry in the QM model
eliminates the possibility to address confinement within this model.
However, as recently shown, the confining properties of QCD can be
effectively accounted for by introducing the Polyakov loop in the
chiral quark--meson model~\cite {Schaefer:2007pw}, in complete analogy
to the PNJL model~\cite{fukushima}. Consequently, the PQM model is an
effective model of both the chiral and confining properties of QCD.

The Lagrangian of the  PQM model  reads
\begin{eqnarray}\label{eq:pqm_lagrangian}
  {\cal L} &=& \bar{q} \, \left[i \gamma^\mu D_\mu - g (\sigma + i \gamma_5
  \vec \tau \cdot \vec \pi )\right]\,q
  +\frac 1 2 (\partial_\mu \sigma)^2+ \frac{ 1}{2}
  (\partial_\mu \vec \pi)^2
  \nonumber \\
  &&  - U(\sigma, \vec \pi )  -{\cal U}(\ell,\ell^{*})\ ,
\end{eqnarray}
where ${\cal U}(\ell,\ell^{*})$ is an effective potential of the gluon
field expressed in terms of the thermal expectation values of the
color trace of the Polyakov loop and its conjugate,

\begin{equation}
\ell=\frac{1}{N_c}\vev{\Tr_c L(\vec{x})},\quad \ell^{*}=\frac{1}{N_c}\vev{\Tr_c
L^{\dagger}(\vec{x})},
\end{equation}
with
\begin{eqnarray}
   L(\vec x)={\mathcal P} \exp \left( i \int_0^\beta d\tau A_4(\vec
x , \tau)
  \right)\,.
\end{eqnarray}
Here ${\mathcal P}$ stands for the path ordering, $\beta=1/T$ and
$A_4=iA_0$. We use the short-hand notation $A_\mu = g \lambda^a
A^a_\mu/2$.

The $O(4)$ representation of the meson fields is
$\phi=(\sigma,\vec{\pi})$ and the corresponding $SU(2)_L\otimes
SU(2)_R$ chiral representation is given by $\sigma+i \gamma_5
\vec{\tau}\cdot\vec{\pi}$. This implies that there are $N_f^2=4$
mesonic degrees of freedom coupled to $N_f=2$ flavors of constituent
quarks. The coupling between the effective gluon field and quarks is
implemented through the covariant derivative
\begin{equation}
 D_{\mu}=\del_{\mu}-iA_{\mu},
\end{equation}
with $A_{\mu}=\delta_{\mu0}A_0$.
The purely mesonic potential of the model reads
\begin{equation}
U(\sigma,\vec{\pi})=\frac{\lambda}{4}\left(\sigma^2+\vec{\pi}^2-v^2\right)^2-h\sigma,
\end{equation}
while the Polyakov loop potential  is parameterized
by a $Z(3)$ invariant form:
\begin{equation}
 \frac{{\cal U}(T;\ell,\ell^{*})}{T^4}=
-\frac{b_2(T)}{2}\ell^{*}\ell
-\frac{b_3}{6}(\ell^3 + \ell^{*3})
+\frac{b_4}{4}(\ell^{*}\ell)^2\,\label{eff_potential}.
\end{equation}
The parameters,
\begin{eqnarray}
\hspace{-4ex}
  b_2(T) &=& a_0  + a_1 \left(\frac{T_0}{T}\right) + a_2
  \left(\frac{T_0}{T}\right)^2 + a_3 \left(\frac{T_0}{T}\right)^3\,,
\end{eqnarray}
where $a_0 = 6.75$, $a_1 = -1.95$, $a_2 = 2.625$, $a_3 = -7.44$, $b_3
= 0.75$ and $b_4 = 7.5$ are chosen to reproduce the equation of state
obtained on the lattice for pure SU(3) gauge theory~\cite{PNJL}, which
yields a first-order phase transition at the temperature $T_0=270$
MeV.

\begin{widetext}

  The thermodynamic potential of the PQM model is in the mean-field
  approximation~\cite{Schaefer:2007pw}

\begin{equation}
  \Omega_{\rm MF}(T,\mu;\langle\sigma\rangle,\ell,\ell^*)  = {\cal
    U}(T;\ell,\ell^*) + U(\langle\sigma\rangle, \vev{\vec{\pi}}=\vec{0}\,) +
  \Omega_{q\bar{q}} (T,\mu;\langle\sigma\rangle,\ell,\ell^*). 
\label{Omega_MF}
\end{equation}
The quark contribution, with dynamical mass
$m_q=g\langle\sigma\rangle$ and energy $E_q =
\sqrt{\vec{p}^{\,2}+m_q^2}$, is given by
\begin{equation}
\Omega_{q\bar{q}} (T,\mu;\langle\sigma\rangle, \ell,\ell^*) = - 2 N_f  \int
\frac{d^3 p}{(2\pi)^3} \left\{ 
 {N_c E_q} \theta( \Lambda^2 - \vec{p}^{\,2})  + T \ln
 g^{(+)}(T,\mu; \langle\sigma\rangle, \ell, \ell^*) +  T \ln
 g^{(-)}(T,\mu; \langle\sigma\rangle, \ell, \ell^*) \right\}, 
\label{Omega_MF_q}
\end{equation}
where
\begin{eqnarray}
g^{(+)}(T,\mu; \langle\sigma\rangle,\ell, \ell^*) &=& 1 + 3 \ell
e^{-(E_q-\mu)/T} + 3 \ell^*e^{-2(E_q-\mu)/T} + e^{-3(E_q-\mu)/T}, \\ 
g^{(-)}(T,\mu; \langle\sigma\rangle,\ell, \ell^*) &=& g^{(+)}
(T,-\mu;\langle\sigma\rangle, \ell^*, \ell)\ . 
\label{g}
\end{eqnarray}
\end{widetext}
The first term in Eq.~(\ref{Omega_MF_q}) is the fermion vacuum
contribution, regularized by the ultraviolet cutoff $\Lambda$.

The equations of motion for the mean fields $\langle\sigma\rangle$,
$\ell$ and $\ell^*$ follow from the stationarity conditions
\begin{equation}
  \frac{\partial \Omega_{\rm MF}}{\partial
      \sigma} =
  \frac{\partial \Omega_{\rm MF}}{\partial \ell} = \frac{\partial
    \Omega_{\rm MF}}{\partial \ell^*} =0\ ,
\label{EoMMF}
\end{equation}
which yield the temperature  and chemical potential dependence of the chiral $\vev{\sigma}
(T,\mu)$ as well as of the Polyakov loops  $\ell
(T,\mu)$ and $\ell^*(T,\mu)$  order parameters.

The parameters $h$, $g$, $\lambda$ and $v$ are specified by requiring
that the following vacuum properties be reproduced: the chiral
condensate is fixed to $\vev{\sigma}(0,0) = f_\pi$, the pion mass to
$m_\pi^2=h/f_\pi$, the constituent mass of quarks to $m_{q}^{0} = g
f_\pi$ and the sigma mass to
\begin{equation}
m_\sigma ^2 = \frac{\partial^2 \Omega_{\rm MF}(0,0; f_\pi,\ell,\ell^*)}{\partial \sigma^2}
\label{msigma2}
\end{equation}
at the  potential minimum.
In the following we use the values $m_\pi=138$ MeV, $m_\sigma=700$
MeV, $m_{q}^{0}=335$ MeV and $f_\pi=93$ MeV.

The results in the chiral limit are obtained with the same parameters
except for the symmetry breaking term $h$, which is set to zero. Thus,
we ignore any influence of the pion mass on the other parameters, like
e.g. $f_\pi$ ~\cite{Gasser:1984gg}. For the problem at hand, such
shifts are unimportant.

The thermodynamic potential in the QM model is obtained from
Eq.~(\ref{Omega_MF}) by setting $\ell=\ell^*=1$ and ignoring the
constant contribution from the Polyakov loop potential. In this case
the functions $g^{(\pm)}$ reduce to
\begin{equation}
g^{(\pm)}(T,\mu;\langle\sigma\rangle,\ell=1,\ell^*=1) = \left[1 -
  n_F(T,\pm\mu;\langle\sigma\rangle)\right]^{-3}, \nonumber 
\end{equation}
where
\begin{equation}
n_F (T,\pm\mu;\langle\sigma\rangle) = \left[\exp\left(
    \frac{E_q\pm\mu}{T}  \right) +1 \right]^{-1} 
\label{FD}
\end{equation}
is the Fermi-Dirac distribution function. In the following, we do not
consider fluctuations of the $\sigma$ field. Hence, we simplify the
notation by dropping the bra-ket notation. Thus, from this point on,
$\sigma\equiv\langle\sigma\rangle$ denotes a uniform classical field.

\begin{figure*}
\includegraphics*[width=7.3cm]{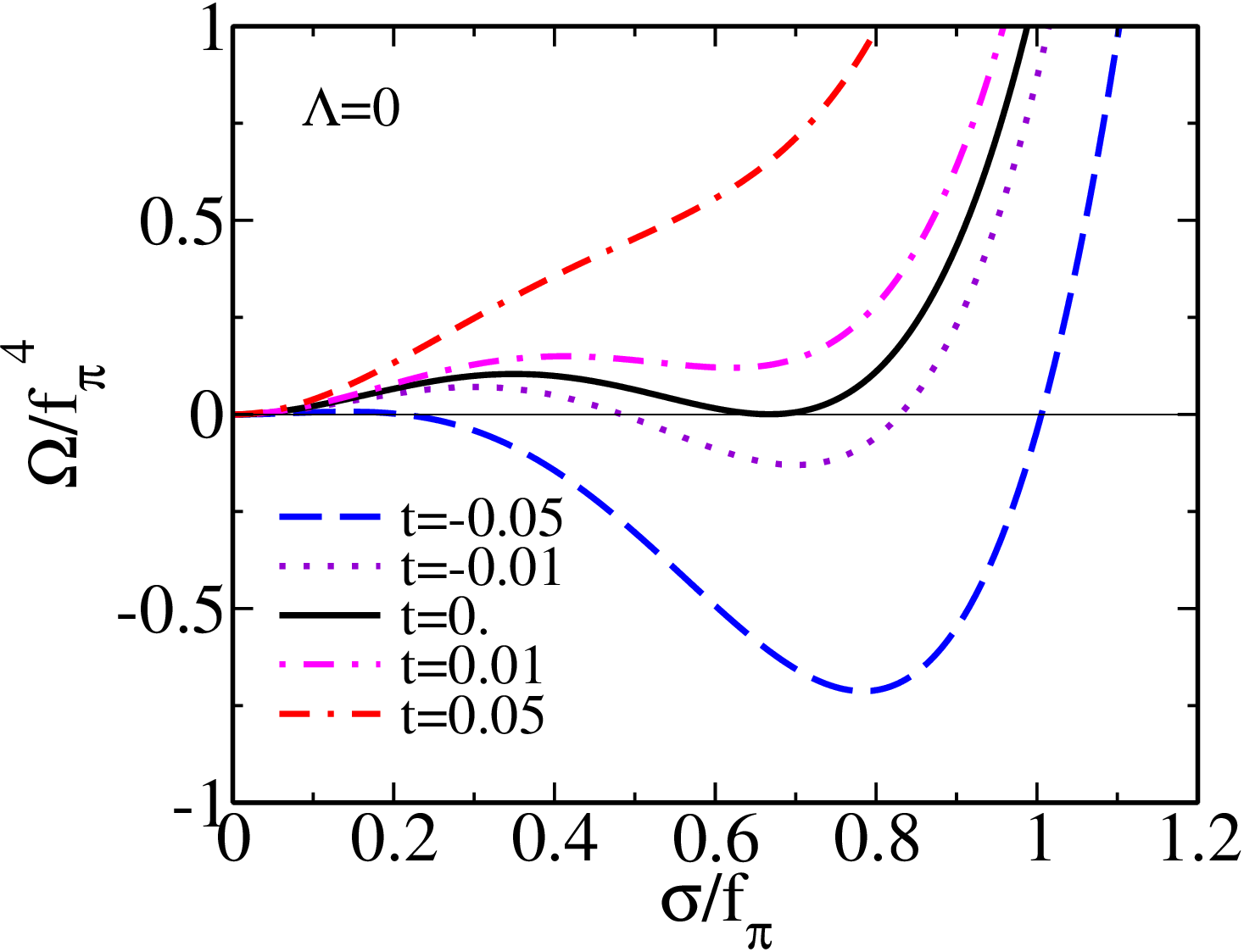}
\includegraphics*[width=7.5cm]{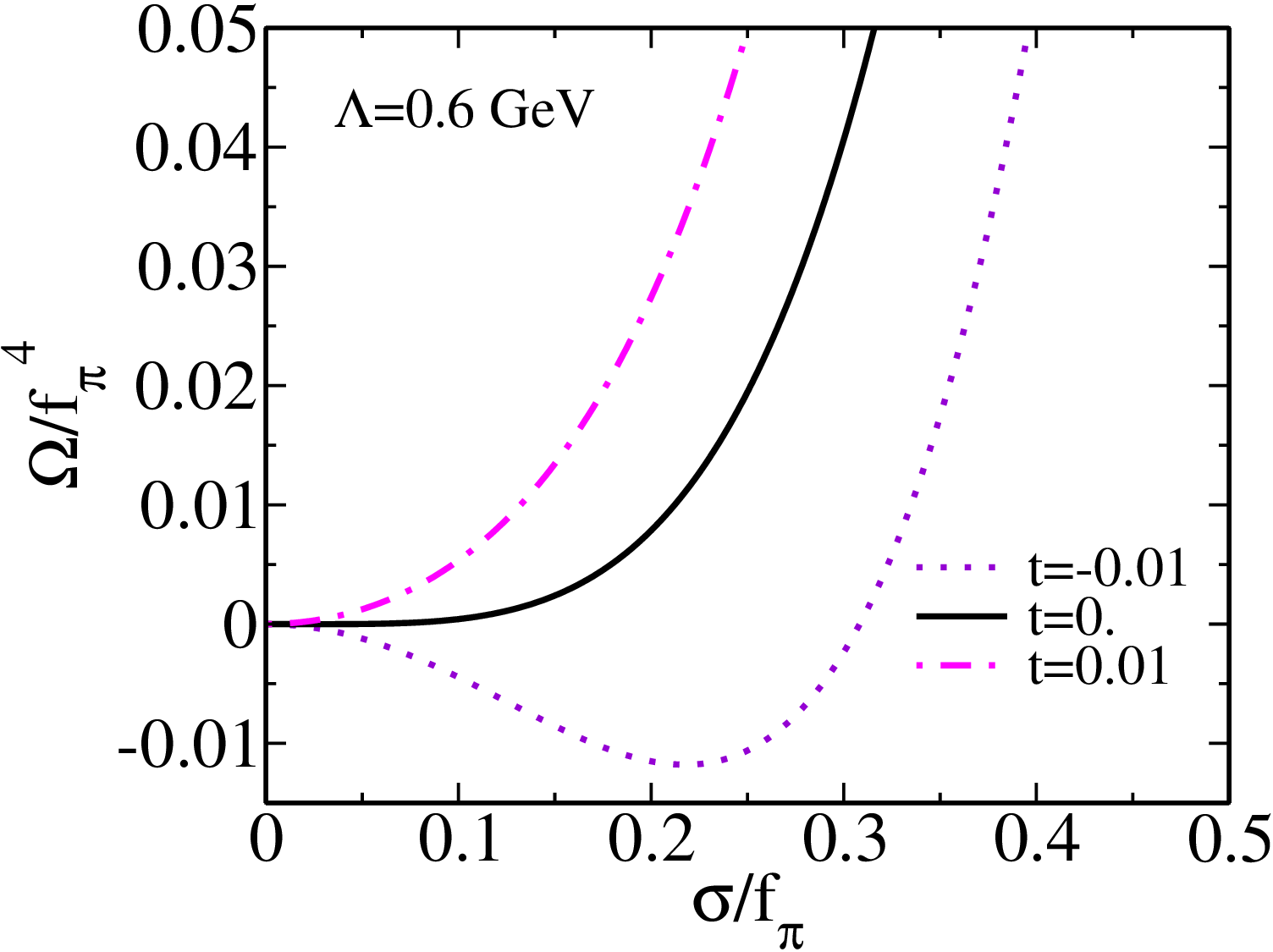}
\caption{ The effective potential as a function of the order parameter
  in the vicinity of the phase transition for vanishing (left panel)
  and finite (right panel) cutoffs. The curves correspond to $\mu=0$
  and a few temperature close to $T_{PT}$. The potential is normalized
  to zero at the origin. }
\label{effec_poten}
\end{figure*}

\section{Landau effective theory}

The mean-field thermodynamic potential (\ref{Omega_MF}) can be used to
explore the influence of fermion vacuum fluctuations on the critical
and thermal properties of a system of quarks interacting with the
meson and gluon mean fields. In this section we explore the role of
the fermion vacuum term in an effective Landau theory for the QM model
with and without the fermion vacuum term. Thus, for the moment we
neglect the Polyakov loop potential and its coupling to the quarks.
The conclusions of this section are, however, directly applicable also
to the Polyakov loop extension of the QM model.

\subsection{Effective potential without  fermion  vacuum fluctuations }

The contribution of the fermion vacuum term to the effective
potential~(\ref{Omega_MF_q}) can be eliminated by setting the cutoff
parameter $\Lambda$ to zero. In the chiral limit, i.e., for $h\to 0$,
and near the chiral phase transition the dynamical quark mass, which
is proportional to the order parameter, is very small. Consequently,
the thermodynamics of the QM model can be formulated as an effective
theory for the order parameter. Due to the smallness of the quark mass
the high temperature expansion can be applied to derive the effective
quark potential ~\cite{Dolan:1973qd, Quiros:1999jp, Landsman:1986uw}
\begin{eqnarray}
\label{thermal}
&&\Omega_{q\bar{q}}^{\rm th} \simeq
N_c N_f T^4\left\{
-\frac{7 \pi^2}{180}  +  \frac{1}{12} \frac{m_q^2}{T^2} +
\frac{1}{8\pi^2}\frac{m_q^4}{T^4}\times\right. \\ &&\left. \left[ 
\ln\left( \frac{m_q}{\pi T} \right) + \gamma_E - \frac34
\right]
+ \frac{7\zeta(3)}{192\pi^4} \frac{m_q^6}{T^6}
 + \mathcal{O}\left( \frac{m_q^8}{T^8}\right)\right\}, \nonumber
\end{eqnarray}
where $\gamma_E$ is Euler's constant. Note that there is no cubic
quark mass term in this expansion in contrast to the bosonic case.

The effective potential of the QM model is the sum of the quark and
meson contributions
\begin{equation}
\Omega(\sigma) = \Omega_{q\bar{q}}^{\rm th}  - \frac{\lambda v^2}{2}\sigma^2+\frac{\lambda}{4}\sigma^4.
\label{OmegaHT}
\end{equation}
We note that for $\lambda > 0$, one would expect the phase transition to be second order.

Because the dynamical quark mass is proportional to the order
parameter $\sigma$, the logarithmic term in Eq. (\ref{thermal})
competes with the quartic term in the meson potential (\ref{OmegaHT}).
For sufficiently small values of the order parameter, the logarithmic
term is arbitrarily large and negative. This leads to a change in sign
of the effective quartic coupling in Eq. (\ref{OmegaHT}), which
implies that the phase transition is first order for any finite value
of $\lambda$. This effect resembles the Coleman-Weinberg
fluctuation-induced first-order phase transition~\cite{Coleman:1973jx}
at $T=0$. In order to make the mechanism more transparent, we
introduce the Landau effective potential, corresponding to Eq.
(\ref{OmegaHT})
\begin{equation}
\Omega(\sigma) - \Omega_{\rm bg} = \frac{1}{2} a(T) \sigma^2 + \frac14 b \sigma^4 [1+c\ln(\sigma/\sigma_0)],
\label{GL}
\end{equation}
where, to linear order in $T-T_{c}$, $a(T)=A \cdot (T-T_c)$ and $A>0$.
The quartic coupling $b$ as well as the coefficient $c$ are assumed to
be positive constants. In general, both $b$ and $c$ depend on
temperature. However, this dependence is irrelevant for our
discussion, as long as $b(T_c)$ and $c(T_c)$ are non-zero.

The parameters of the Landau potential (\ref{GL}) can be related to
that of the QM model. By inspection of Eq.~(\ref{thermal}) and
(\ref{OmegaHT}) one finds
\begin{eqnarray}
a &=& \frac{g^2 N_c N_f}{6} (T^2-T_c^2)\nonumber\\
&=&  A \cdot  (T-T_c) +\mathcal{O}\left[\left(T-T_{c}\right)^{2}\right],
\label{aHT}
\end{eqnarray}
where
\begin{eqnarray}
\label{TcHT}
T_c^2 &=&  \frac{6\lambda v^2}{g^2 N_c N_f}, \\
A &=&  g^2 N_c N_f T_c/3.
\label{AHT}
\end{eqnarray}
Similarly, one can identify the quartic coupling $b$ and the
coefficient $c$
\begin{eqnarray}
\label{bHT}
b &=& \lambda + \frac{g^4 N_c N_f} { 2 \pi^2} \left(\gamma_E - \frac 34 \right), \\
c &=& \frac{g^4 N_c N_f} { 2\pi^2 b},
\label{cHT}
\end{eqnarray}
as well as the parameter $\sigma_0$, which for $T\simeq T_c$ is given
by
\begin{equation}
\sigma_0 = \frac{\pi T}{g}\approx  \frac{\pi T_c}{g}.
\label{sigma0HT}
\end{equation}
The background contribution $\Omega_{\rm bg}$ in Eq. (\ref{GL}) is
independent of the order parameter $\sigma$, and given by the $T^4$
term in Eq. (\ref{thermal}).

For small positive values of $a$, i.e. for $T\gtrsim T_c$, the
potential (\ref{GL}) has two minima located at
\begin{eqnarray}
\sigma_1/\sigma_0 &=& 0,\\
\sigma_2/\sigma_0  &=& \kappa - \frac{a}{2bc\kappa\sigma_0^2},
\label{minima}
\end{eqnarray}
where $\kappa=\exp(-1/4-1/c)$.
At these points, the corresponding values of the potential are
\begin{eqnarray}
&&\Omega(\sigma_1)- \Omega_{\rm bg}= 0, \\
&&\Omega(\sigma_2)- \Omega_{\rm bg}= - \frac14 b c (\sigma_0 \kappa)^4 + a (\sigma_0 \kappa)^2.
\label{values}
\end{eqnarray}
Consequently, for $a<\frac14 bc (\sigma_0 \kappa)^2$ the second
solution, $\sigma_2$, is energetically favored. For $a=\frac14 bc
(\sigma_0 \kappa)^2$ the two minima at $\sigma_1=0$ and
$\sigma_2=\frac34 \kappa \sigma_0$ are degenerate, i.e.
$\Omega(\sigma_1)=\Omega(\sigma_2)=\Omega_{\rm bg}$. Finally, for
$a>\frac14 bc (\sigma_0 \kappa)^2$ the first solution $\sigma_1$ is
energetically favored. This analysis indicates that a first-order
phase transition occurs at $T_{PT} = T_c + bc (\kappa \sigma_0)^2/(4A)
$, i.e. at a temperature above the nominal critical temperature $T_c$.

The results obtained above are illustrated in
Fig.~\ref{effec_poten}-left, where we show the effective potential of
the QM model for $\mu=0$, neglecting the fermion vacuum term. The
dependence of the potential on the sigma field, for selected values of
the reduced temperature $t=(T-T_{PT})/T_{PT}$, shows explicitly that
this model exhibits a first-order phase transition in the chiral
limit~\footnote{Note that here the transition temperature $T_{PT}$ is
  not given by the analytical expression given above, but is
  determined numerically within the QM model.}. Thus, without the
fermion vacuum term, this model does not reproduce the
second-order transition of QCD with two massless flavors,
expected from the universality argument.

\subsection{Effective potential with the fermion vacuum contribution }
We now include the contribution of the fermion vacuum term of the QM
model and analyze the resulting thermodynamic potential. For
sufficiently small values of the dynamical quark mass $m_q$, the
vacuum term in Eq.~ (\ref{Omega_MF_q}) can be expanded in a Puiseux
series in $m_q/\Lambda$. The leading terms of this expansion reads
\begin{eqnarray}
\label{expansion}
&&\Omega_{q\bar{q}}^{\rm vac} =  - 2 N_f N_c  \int \frac{d^3
  p}{(2\pi)^3} E_q \theta(\Lambda^2 - \vec{p}^{\,2}  )  \simeq \\
\nonumber 
&& - \frac{N_f N_c\Lambda^4}{4 \pi^2}  \left\{ 1 +
  \frac{m_q^2}{\Lambda^2}  + \frac{1}{8}\frac{m_q^4}{\Lambda^4}
  \left[1 + 4 \ln\left( \frac{m_q}{2\Lambda}\right)\right] \right.
\nonumber \\ &&- \left.\frac{1} 
  {8} \frac{m_q^6}{\Lambda^6} +
  \mathcal{O}\left(\frac{m_q^8}{\Lambda^8}\right)\right\}. \nonumber 
\end{eqnarray}
The logarithmic dependence on $m_q$ cancels between the thermal and
vacuum contributions to the thermodynamic potential, (\ref{thermal})
and (\ref{expansion}). Consequently, when the fermion vacuum term is
included, the effective potential has the typical structure expected
for a theory with the second-order phase transition
\begin{equation}
\Omega(\sigma) - \Omega_{\rm bg} = \frac{1}{2} a(T) \sigma^2 + \frac14 b \sigma^4 .
\label{GLwithVT}
\end{equation}
The mass coefficient is formally given by the Eq.~(\ref{aHT}), however
with the following redefinition of the critical temperature
\begin{equation}
T_c^2 = \frac{3}{\pi^2} \Lambda^2 +  \frac{6 \lambda v^2}{g^2 N_c N_f}
\label{TcL}
\end{equation}
and the quartic coupling reads
\begin{equation}
b = \lambda + \frac{g^4 N_c N_f}{2 \pi^2} \left[\gamma_E -1 +
  \ln\left(  \frac{2\Lambda}{\pi T_c} \right) \right]. 
\label{bL}
\end{equation}

In the above expressions the parameters $\lambda$ and $v^2$ depend on
the cutoff $\Lambda$, because we require that the physical vacuum
parameters discussed above should be reproduced independently of the
cutoff. Taking this dependence into account, the leading $\Lambda^2$
term in Eq.~(\ref{TcL}) is canceled by a corresponding term in
$\lambda v^2$. Consequently, for large $\Lambda$, the critical
temperature is cutoff independent:
\begin{equation}
T_c^2(\Lambda\gg T_c) = \frac{3 m_\sigma^2 f_\pi^2} {N_c N_f  (m_{q}^0)^2} +  \frac{3(m_{q}^0)^2}{2\pi^2} + \mathcal{O}(\Lambda^{-2}).
\label{TcLInf}
\end{equation}

The quartic coupling is not positive definite for all values of $\Lambda$.
For $\Lambda>\Lambda^*$, where $\Lambda^*$ is defined by $b(T\to T_c, \Lambda^*)=0$, the thermodynamic potential (\ref{GLwithVT}) has only one minimum at any temperature,  at $\sigma_1=0$ or at $\sigma_2=\sqrt{-a/b}$. Thus, 
a second-order phase transition takes place at $T_{PT}=T_c$. This behavior is illustrated 
 in Fig.~\ref{effec_poten}-right, where we show the thermodynamic potential obtained in the QM model, including the fermion vacuum term,  for  $\Lambda=0.6$ GeV, which implies $\lambda = 3.06$ and $v^2 = - (578)^2$ MeV$^2$. For this choice of parameters, the position of the minimum of the thermodynamic potential varies continuously with temperature,  indicating a second-order phase transition.

\begin{figure}
\includegraphics*[width=7cm]{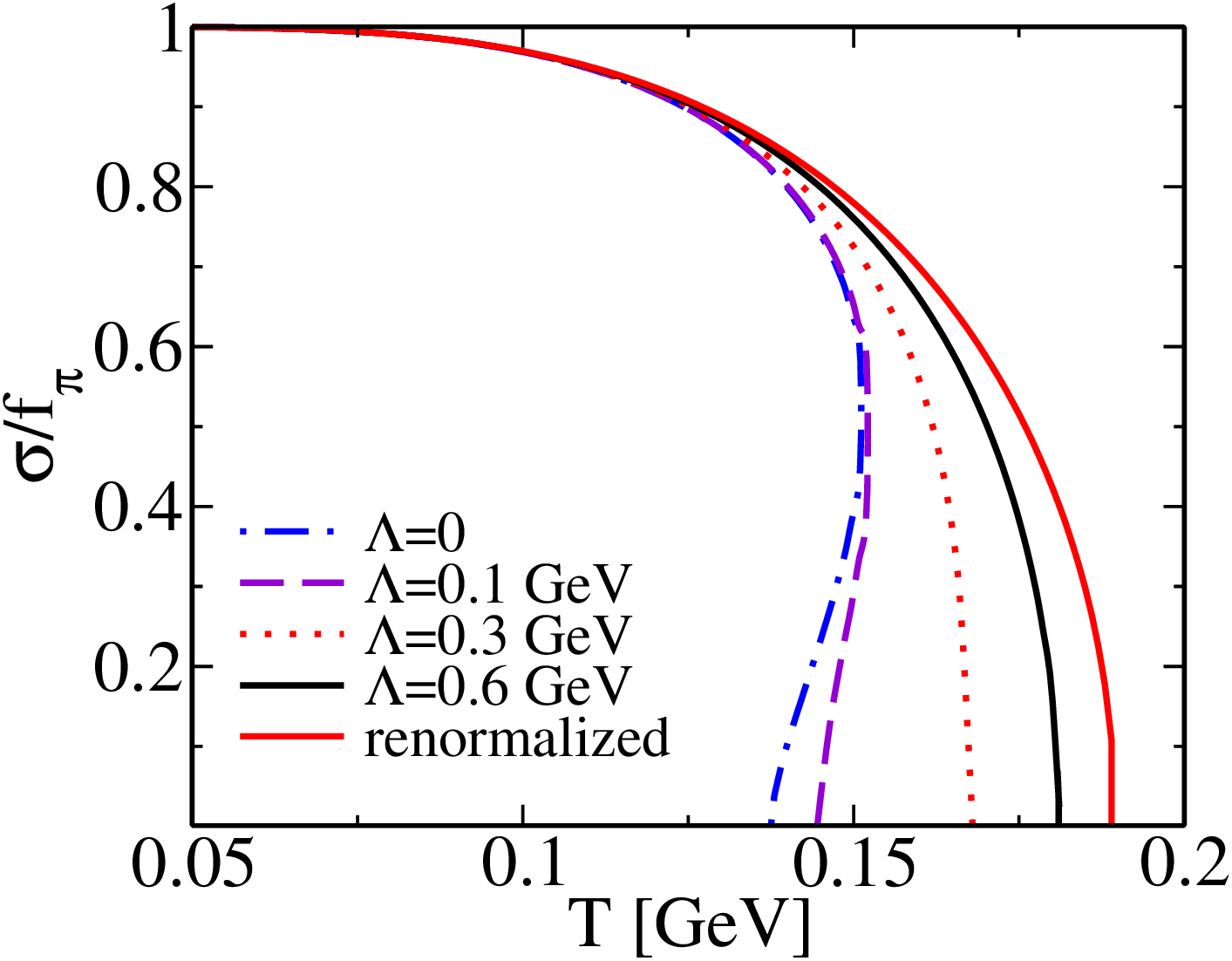}
\caption{
The order parameter as a function of temperature for different values of the cutoff parameters. The uppermost line corresponds to the  
renormalized theory.
}
\label{sigma}
\end{figure}

However, for $\Lambda<\Lambda^*$, the quartic coupling is negative,
while, as shown in Eqs.~(\ref{thermal}) and~(\ref{expansion}), the
coefficient of the $\sigma^6$ term is always positive, as required for
stability. The negative value of $b$ results in the appearance of the
second minimum in the potential, leading to the first-order phase
transition. This is similar to the situation described in the previous
section, where the fermion vacuum term was neglected. The appearance
of a first-order phase transition for small $\Lambda$ is illustrated
in Fig.~\ref{sigma}, where the temperature dependence of the order
parameter is shown for several values of the cutoff $\Lambda$. For
$\Lambda<0.3$ GeV, the order parameter exhibits hysteresis,
characteristic of a first-order phase transition. For the parameter
set employed here, we find $\Lambda^*\simeq 244$ MeV. This value is
comparable to the physical scales that characterizes the vacuum. In
general, the cutoff should be larger than any physical scale of a
system. In this case we find a second-order chiral phase transition in
the QM model.

The fermion vacuum fluctuations modify the critical properties of the
QM model considerably. The modification depends on the value of the
ultraviolet cutoff $\Lambda$, which is introduced to regularize the
divergent vacuum term. For large cutoffs ($\Lambda>\Lambda^*$), the QM
model exhibits a second-order phase transition, whereas in the
opposite limit the phase transition stays first order, as in the
approximation, where the vacuum term is completely neglected.

Clearly, a finite cutoff modifies not only the critical structure of
the medium but also the properties of physical observables. Hence, it
is desirable to remove the cutoff dependence by means of a suitable
renormalization procedure. This problem will be addressed in the next
section.

\section{Renormalization  of the fermion vacuum term}
The expansion (\ref{expansion}) indicates that, in addition to the
fermion vacuum term, also the mass parameter $a$ and the quartic
coupling $b$ diverge in the limit of $\Lambda\to\infty$. These
divergencies can be regularized by introducing a cutoff, as done
above. In this section we choose a more convenient route and use the
dimensional regularization scheme.

The vacuum term is, to lowest order, just the one-loop effective
potential at zero temperature~\cite{Quiros:1999jp}
\begin{eqnarray}
\label{vt_one_loop}
\Omega_{q\bar{q}}^{\rm vac} &=&   - 2 N_f N_c  \int \frac{d^3
  p}{(2\pi)^3} E_q  \nonumber\\ 
&=&    - 2 N_f N_c  \int \frac{d^4 p}{(2\pi)^4}  \ln(p_0^2+E_{q}^{2})
+ {\rm C}, 
\end{eqnarray}
where the infinite constant $C$ is dropped, since it is independent of
the fermion mass.

To regularize Eq.~(\ref{vt_one_loop}) we perform the dimensional
regularization near three dimensions, $d=3-2\epsilon$. The resulting
potential up to zeroth order in $\epsilon$ reads
\begin{equation}
\Omega_{q\bar{q}}^{\rm vac} =  \frac{N_c N_f}{16 \pi^2} m_q^4 \left\{
  \frac{1}{\epsilon} - \frac{1}{2} 
\left[  -3 + 2 \gamma_E + 4 \ln\left(\frac{m_q}{2\sqrt{\pi} M}\right)
\right] 
\right\},
\label{Omega_DR}
\end{equation}
where $M$ is an arbitrary  renormalization scale parameter.

The thermodynamic potential is then renormalized by adding a counter
term to the Lagrangian of the QM or PQM model. With the
choice~\footnote{ Here we use the freedom in the definition of counter
  terms to remove irrelevant constants.}
\begin{equation}
\delta \mathcal{L} = \frac{N_c N_f}{16 \pi^2} g^4\sigma^{4} \left\{ \frac{1}{\epsilon} - \frac{1}{2}
\left[  -3 + 2 \gamma_E - 4 \ln\left(2\sqrt{\pi}\right)  \right] \right\},
\label{counter}
\end{equation}
the renormalized contribution of the fermion vacuum loop reads
\begin{equation}
\Omega_{q\bar{q}}^{\rm reg} =  -\frac{N_c N_f}{8 \pi^2} m_q^4  \ln\left(\frac{m_q}{ M}\right).
\label{Omega_reg}
\end{equation}
The quark contribution to the thermodynamic potential is then given by
Eq.~(\ref{Omega_MF_q}), with the vacuum term replaced by
Eq.~(\ref{Omega_reg}).

For large values of the order parameter, $m_q = g \sigma > M$, the
renormalized vacuum contribution in Eq.~(\ref{Omega_reg}) is unbounded
from below. At $T=0$, the potential (\ref{Omega_MF}), including the
renormalized quark vacuum term, has a minimum at the physical point
$\sigma=f_\pi$, a maximum at $\sigma\simeq 212$ MeV and then decreases
monotonously to $-\infty$ for $\sigma\to \infty$ \footnote{We note
  that for finite $\Lambda$, the cutoff regularized
  potential, Eq.~(\ref{expansion}), is always bounded from below.}. The
instability for large $\sigma$ is symptomatic of the renormalized
one-loop approximation~\cite {Coleman:1973jx}.
The inclusion of higher order loop contributions is known to cure this
problem.

Dimensional regularization introduces an arbitrary renormalization
scale parameter $M$. Consequently, the parameters $v$ and $\lambda$
are $M$ dependent through the conditions (\ref{msigma2})
and~(\ref{EoMMF}). Nevertheless, in the one-loop approximation,
physical quantities are independent of $M$~\cite{Coleman:1973jx}.
Indeed, it is rather straightforward to show that
\begin{equation}
\frac{d \Omega_{\rm MF}(\lambda[M], v^2[M],M)} {d M} =0\ .
\label{indepM}
\end{equation} 

Sufficiently close to the phase transition, the Landau potential
corresponding to the renormalized QM model has the
form~(\ref{GLwithVT}), with the mass parameter given by
Eqs.~(\ref{aHT}-\ref{AHT}). The critical temperature $T_c$ is
independent of the renormalization scale, since $\partial (\lambda
v^2)/\partial M =0$. Consequently, it can be expressed solely in terms
of physical quantities,
\begin{equation}
T_c^2 = \frac{3 m_\sigma^2 f_\pi^2} {N_c N_f (m_{q}^{0})^2} +  \frac{3(m_{q}^{0})^2}{2\pi^2}.
\label{TcRe} 
\end{equation}
This expression for $T_{c}$ agrees with the one obtained in the cutoff
regularization scheme, Eq.~(\ref{TcLInf}), for $\Lambda\to\infty$. Also
the quartic coupling $b$ can be expressed in terms of physical
quantities
\begin{equation}
b = \frac{m^2_\sigma}{2 f^2_\pi}  + \frac{g^4 N_c N_f} {2\pi^2} \left[
  \gamma_E + \ln\left(\frac{g f_\pi}{\pi T_c}\right) \right]. 
\label{bRm}
\end{equation}

Thus, we conclude that in the mean-field approximation, the
thermodynamic potential of the quark-meson model, including the
renormalized fermionic vacuum contribution, is uniquely defined. In
Fig.~\ref{sigma} we show the temperature dependence of the order
parameter, as obtained within this model. This and other physical
quantities are, as indicated above, free from ambiguities related to
the choice of the renormalization scale. In the chiral limit, this
model exhibits a second-order phase transition at $\mu=0$, as expected
for two-flavor QCD. Thus, the quark-meson model, with appropriately
{renormalized } fermionic vacuum fluctuations, is an effective
QCD-like model, suitable for describing the physics at the chiral
phase transition.

\begin{figure}
\includegraphics*[width=7cm]{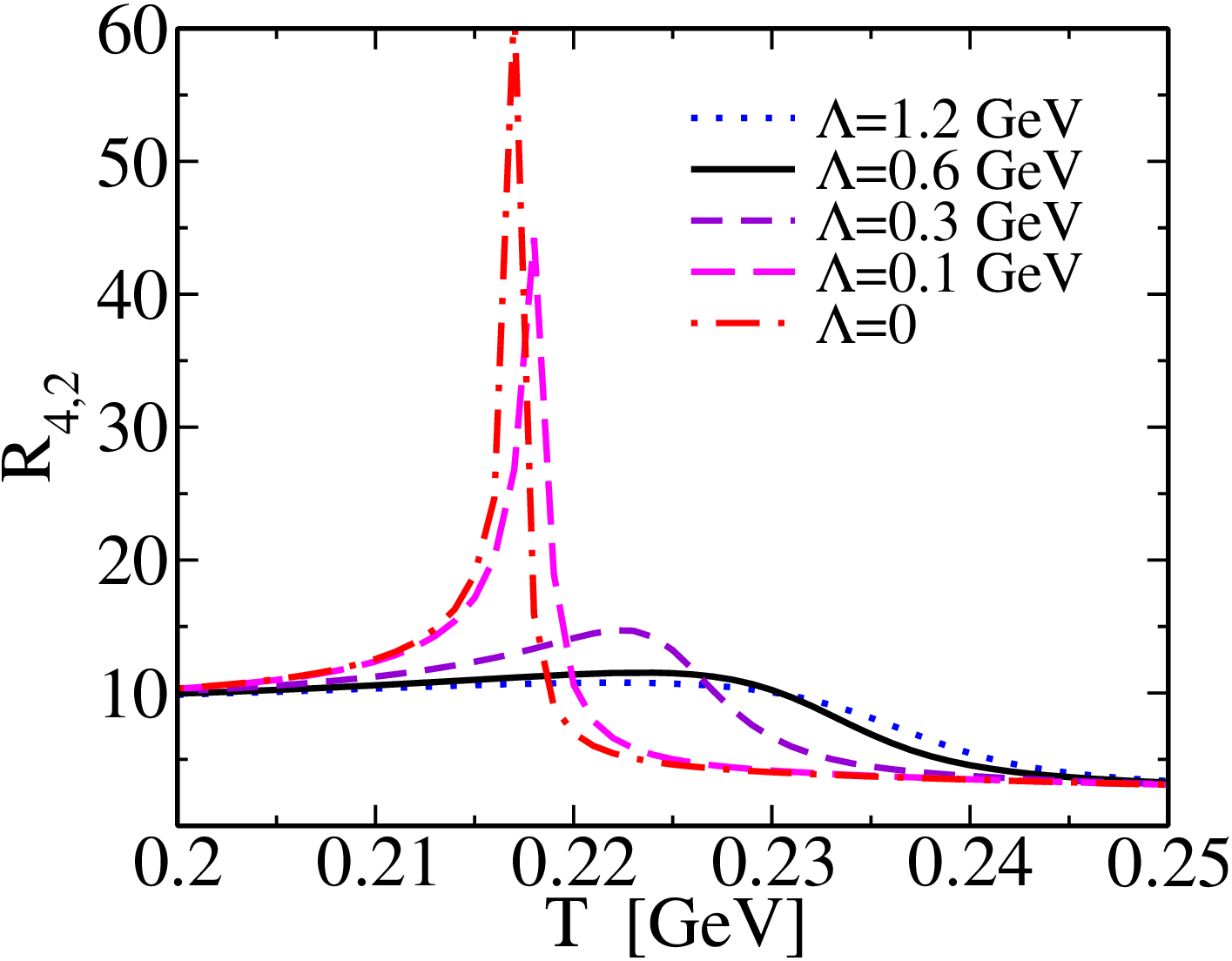}
\caption{ The kurtosis in the PQM model as a function of temperature
  for several values of cutoff parameters $\Lambda$.}
\label{rl}
\end{figure}

\section{Quark number density fluctuations}

As discussed above, the fermion vacuum term is required to obtain the
correct critical behavior in the chiral limit. In this section, we
show that the vacuum term is important also for the thermodynamics for
non-zero pion masses. To illustrate this, we consider fluctuations of
the net quark number density within the PQM model.

The fluctuations of the quark number density are characterized by the
generalized susceptibilities,
\begin{equation}
\left. c_n(T)=\frac{\del^n[p\,(T,\mu)/T^4]}{\del(\mu/T)^n}\right\vert_{\mu=0}.
\end{equation}
The first two coeffients, $c_2$ and $c_4$, are the second and the
fourth cumulants of the net quark number $N_{q}$,
\begin{eqnarray}
c_2 &=& {\frac{\chi_q} {T^2}}=\langle(\delta N_q)^2\rangle,
 \\
c_4 &=& \langle(\delta N_q)^4\rangle-3\langle(\delta N_q)^2\rangle^2. \label{fluctuations}
\end{eqnarray}
Here $\chi_q$ is the quark number susceptibility and $\delta N_q=
N_q-\langle N_q\rangle$.
\begin{figure*}
\includegraphics*[width=7cm]{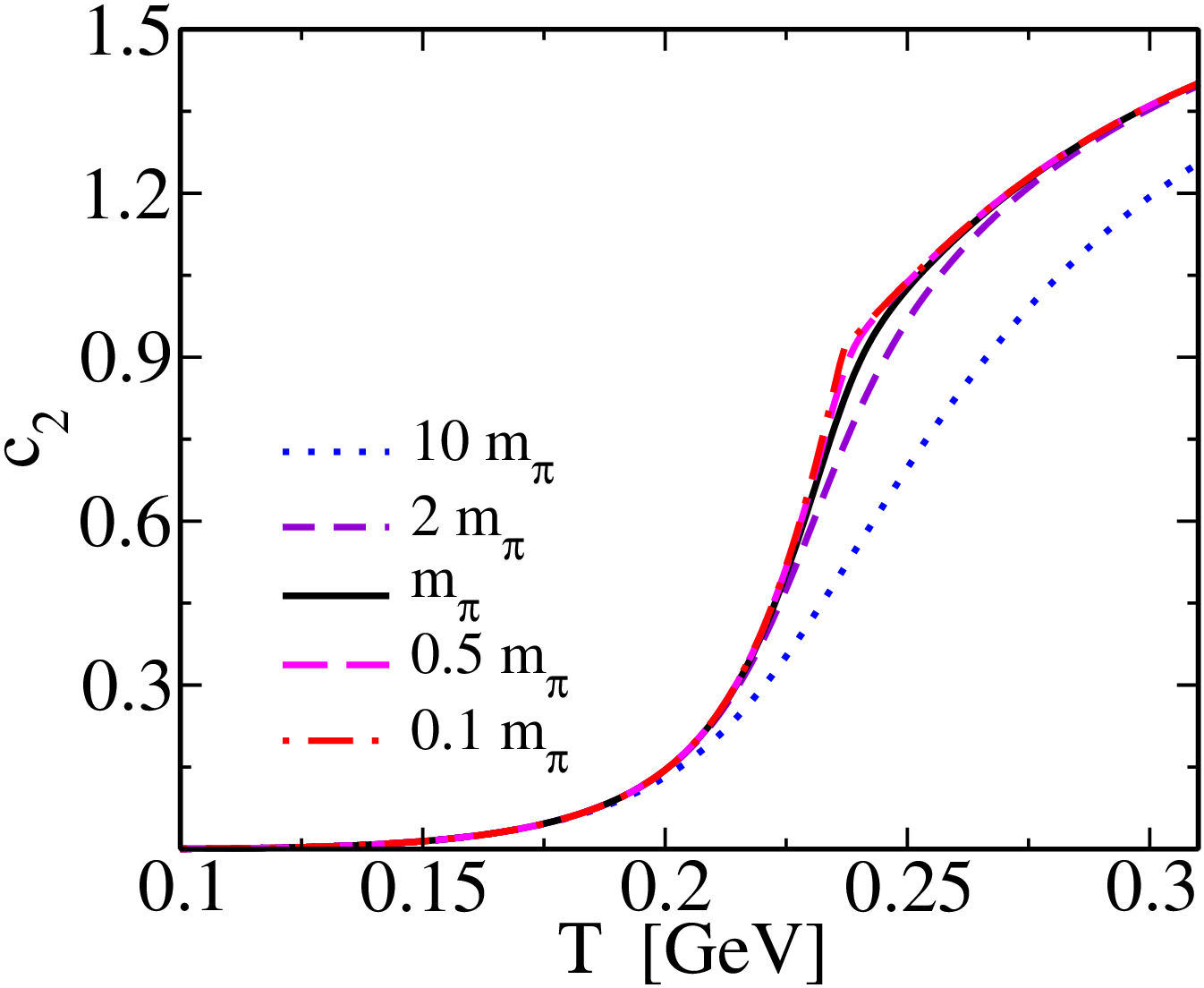}
\includegraphics*[width=7cm]{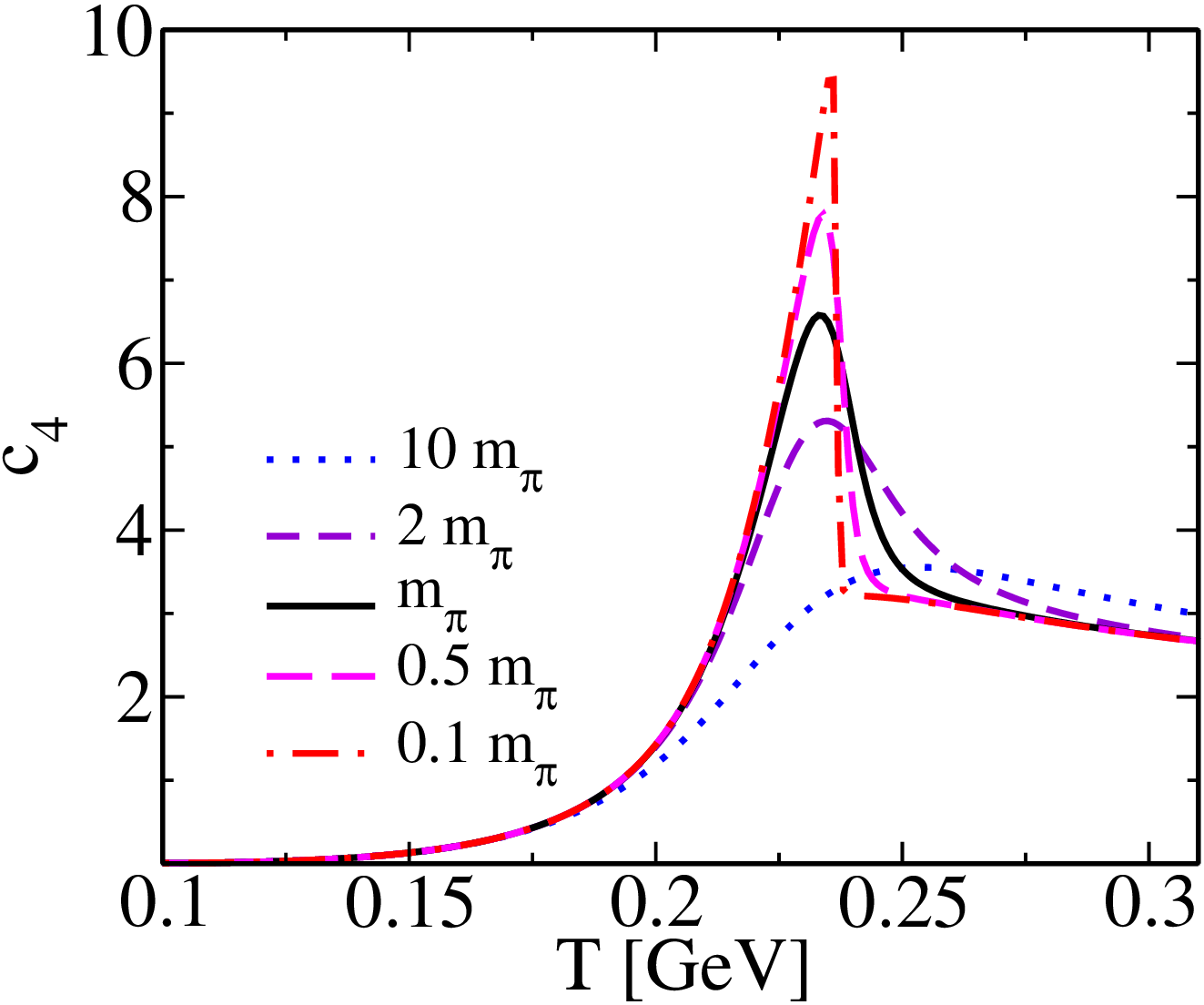}
\caption{ The second $c_2$ and the fourth $c_4$ order cumulants,
  computed in the PQM model including the renormalized fermion vacuum
  potential, as functions of temperature for several values of the
  pion mass. }
\label{c2}
\end{figure*}

The coefficients $c_n(T)$ are sensitive probes of the chiral phase
transition, see e.g. \cite{Wagner:2009pm}. In particular, $c_2$
diverges at the critical endpoint of the QCD phase diagram
\cite{Stephanov:2007fk}. In the chiral limit and at non-zero chemical
potential, all generalized susceptibilities $c_n(T)$ with $n>2$
diverge at the $O(4)$ chiral critical line \cite{Ejiri:2005wq}.
Moreover, if the spinodal instability is reached, fluctuations of the
net quark number also diverge at a first-order chiral phase transition
\cite{cs}.

A very particular role is attributed to the so called kurtosis of the
net quark number fluctuations~\cite{Ejiri:2005wq, F1, kurtosis,
  Schaefer:2009ui}, the ratio of $c_{4}$ and $c_{2}$
\begin{equation}\label{eq:ratio_c42}
 R_{4,2}=\frac{c_4}{c_2}.
\end{equation}
This observable is not only sensitive to the chiral phase transition
but also probes the confinement-deconfinement transition. At very low
and very high temperatures, the kurtosis reflects the quark content of
the baryon-number carrying effective degrees of freedom
~\cite{Ejiri:2005wq, kurtosis}. Thus, at low temperatures, in the
confined phase, $R_{4,2}\simeq N_q^2=9$ while in an ideal gas of
quarks~\footnote{More precisely, this number is $6/\pi^2$ due to
  quantum statistics.} $R_{4,2}\sim 1$. The influence of the chiral
phase transition on the kurtosis has been studied within lattice gauge
theory~\cite{lgt2}. Due to the ``statistical confinement" realized in
the PQM and PNJL models, these models can also be used to explore the
interplay between confinement and chiral restoration e.g. in the
kurtosis~\cite{kurtosis, Schaefer:2009ui}.

We first consider the dependence of the ratio $R_{4,2}$ on the cutoff
$\Lambda$ in the PQM model, including the vacuum contribution,
Eq.~(\ref{expansion}). In Fig.~\ref{rl} the kurtosis is shown at the
physical pion mass, as a function of the temperature for several
values of $\Lambda$. The kurtosis converges to $R_{4,2}=9$ at low
temperatures and to $R_{4,2}\sim 1$ in the high temperature limit for
all values of cutoff parameter. However, near $T_c$ the kurtosis is
very sensitive to $\Lambda$; for small cutoffs it develops a peak,
which is maximal for $\Lambda\to 0$. The sharp peak, obtained for
vanishing cutoff, is a consequence of the underlying first-order phase
transition in the chiral limit. Clearly, such a strong dependence on
the ultraviolet cutoff is unphysical and must be removed by a suitable
renormalization of the vacuum term.

Having established the relevance of the fermion vacuum term for the
kurtosis, we now study fluctuations of the net quark density in the
PQM model, including the renormalized vacuum contribution. In
Fig.~\ref{c2} we show the second and fourth cumulants, $c_2$ and
$c_4$, as functions of temperature for several values of the pion
mass. For small $m_{\pi}$, $c_2$ exhibits a kink at the chiral
crossover transition, while $c_4$ shows an rapid drop, which in the
chiral limit evolves into a discontinuity.

These results can be understood in Landau theory. For small $\sigma$,
the thermodynamic potential reads
\begin{equation}
\Omega(T,\mu) - \Omega_{bg}(T,\mu) = \frac12 a(T,\mu) \sigma^2 +
\frac14 b \sigma^4 -   h \sigma, 
\label{LG}
\end{equation}
where $\Omega_{bg}(T,\mu)$ is the background (regular) contribution,
which is independent of $\sigma$. The external field $h$ breaks the
symmetry and generates a finite pion mass. For a small $\mu$ and $T$
near the chiral transition temperature, the coefficient $a(T,\mu)$ can
be parameterized by~\cite{Ejiri:2005wq}
\begin{equation}
a(T,\mu) = A\frac{T-T_c}{T_{c}}  + B\left(\frac{\mu}{T_{c}}\right)^{2},
\label{a}
\end{equation}
where both, $A$ and $B$ are positive constants. The coefficients of
the effective potential can be expressed in terms of the parameters of
the (P)QM model by considering the expansion of the thermodynamic
potential from Eq.~(\ref{Omega_MF}) for small $m_q$ and $\sigma$.

In the chiral limit and at $\mu=0$ the singular contribution to the
second and fourth cumulants is
\begin{eqnarray}
c_2^{\rm sing} &=&  \frac{A B}{b T_c^{4}} \frac{T-T_c}{T_c}\,
\theta(T_c-T)\ , \\
c_4^{\rm sing} &=& \frac{6 B^2 }{b T_{c}^{4}}\, \theta(T_c-T)\ .
\label{c2c4LG}
\end{eqnarray}
Hence, $c_2$ has a kink at the critical point, while $c_4$ is
discontinuous, in agreement with the numerical results for the PQM
model presented above.

The peak structure in $c_4$ near the chiral transition, shown in
Fig.~\ref{c2}, results from an interplay of the regular part, which
increases as the critical point is approached, and the discontinuity
of the singular part.
\begin{figure}[b]
\includegraphics*[width=7cm]{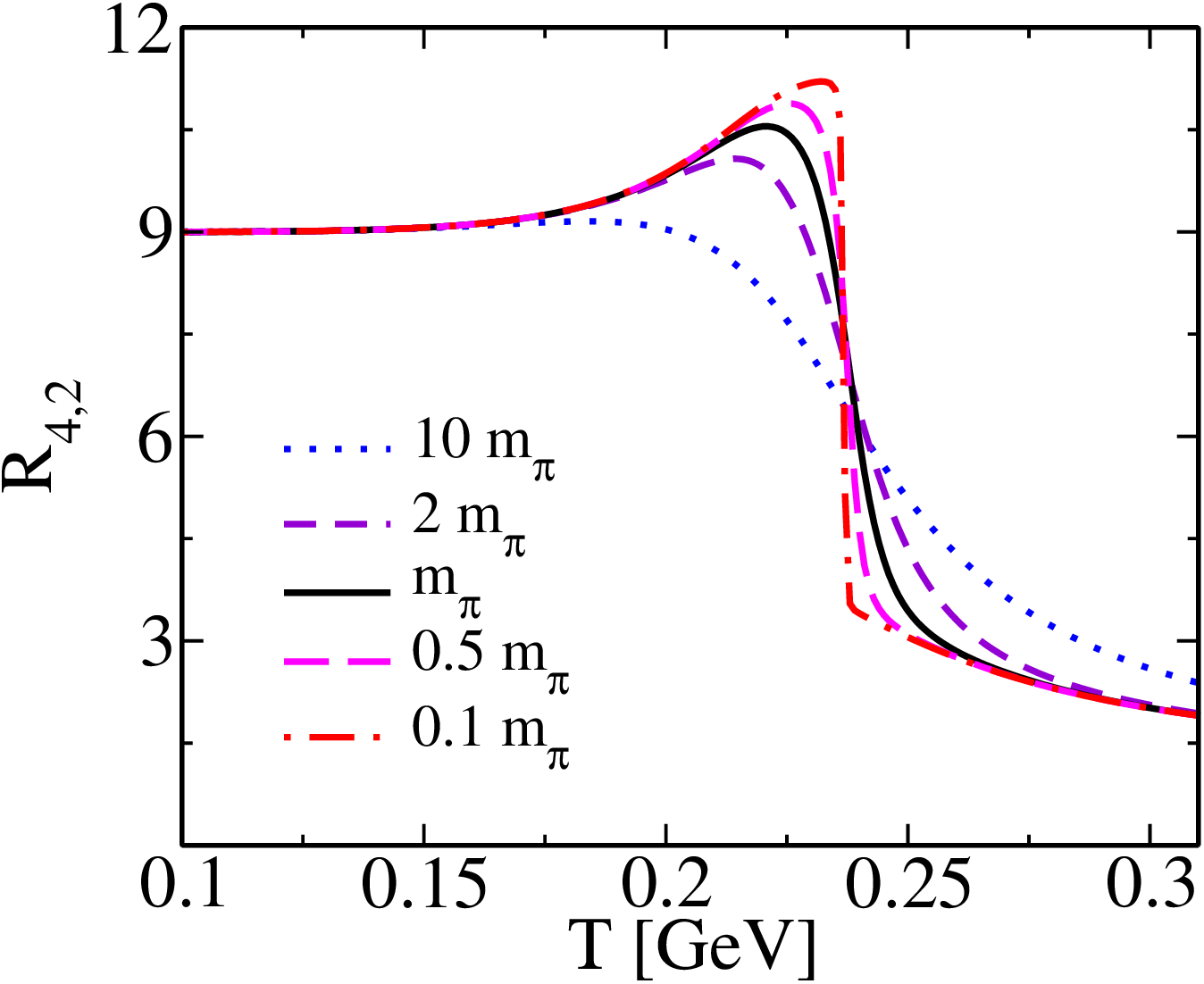}
\caption{ The kurtosis $R_{4,2}$ as a function of temperature for
  several values of the pion mass in the PQM model, including the
  renormalized fermion vacuum term. }
\label{r}
\end{figure}

The dependence of the fourth cumulant $c_4$ on the pion mass is also
reflected in the kurtosis of the quark number fluctuations. In
Fig.~\ref{r} we show $R_{4,2}$ in the vicinity of the chiral phase
transition, for several values of the pion mass. The transition
between the low and high temperature regimes is sharpened as the pion
mass is lowered, indicating that this observable is sensitive not only
to deconfinement but also to chiral dynamics. In the mean-field
approximation, a clear enhancement beyond the low temperature value,
$R_{4,2}=9$, is found just below the pseudo-critical temperature, at
the physical value of the pion mass. There is a further increase of
the enhancement when approaching the chiral limit, where $R_{4,2}$
exhibits a discontinuity at the chiral phase transition, in agreement
with Eq.~(\ref{c2c4LG}). However, when the renormalized vacuum term is
included, this dependence is weaker than previously obtained within
the PQM model, neglecting vacuum fluctuations~\cite{kurtosis}. This
shows, that an appropriate renormalization of the fermionic vacuum
potential is essential for a consistent formulation of the
thermodynamics of the PQM model.

\section{Summary and conclusions}
We have discussed the importance of the divergent fermion vacuum
potential in a consistent formulation of the thermodynamics of the
quark-meson (QM) and the Polyakov loop extended quark-meson (PQM)
models.

We have shown that the vacuum term influences the order of the chiral
phase transition in the chiral limit. The QM and PQM models exhibit a
first-order phase transition at $\mu=0$, when the fermion vacuum term
is neglected. On the other hand, when fermion vacuum fluctuations are
included, a second-order phase transition of the O(4) universality
class is possible, as expected in models that exhibit the $SU(2)\times
SU(2)$ chiral symmetry.

Also for the physical value of the pion mass, the vacuum term
influences thermodynamic observables near the chiral crossover
transition. We have illustrated this explicitly, by computing the
second and the fourth cumulants as well as the kurtosis of the net
quark number fluctuations for several values of the pion mass using
different regularization schemes for the fermion vacuum potential.

Finally, we have renormalized the fermion vacuum term to obtain the
thermodynamic potentials of the QM and PQM models, free of any
dependence on the unphysical ultraviolet cutoff.

\section*{Acknowledgments}
We acknowledge stimulating discussions with Yu.~Ivanov, E.~Kolomeitsev
and V.~Pangon. E.~Nakano and V.~Skokov acknowledge the financial
support by the Frankfurt Institute for Advanced Studies (FIAS).
K.~Redlich acknowledges discussions with H.~Fujii and C.~Sasaki and
partial supports from the Polish Ministry of Science (MEN) and the
Alexander von Humboldt Foundation (AvH).


\begin{thebibliography}{99}

\bibitem{lgt}
A. Bazavov et al.,
Phys. Rev. D {\bf 80}, 014504 (2009);
S. Ejiri et al.,
 Phys.Rev.D {\bf 80},  094505 (2009);
C.R. Allton et al.,
 Phys. Rev. D {\bf 71}, 054508 (2005);
 Y. Aoki et al.,  JHEP {\bf 0906}, 088 (2009);
  Z. Fodor and  S.D. Katz,  Nucl. Phys. News {\bf 16N3}, 12 (2006).

\bibitem{Stephanov:2007fk}
  M.~A.~Stephanov,
  PoS {\bf LAT2006}, 024 (2006).

\bibitem{njl} For a review see e.g., S. P. Klevansky, Rev. Mod. Phys.
  {\bf 64}, 649 (1992).

\bibitem{cs}
C. Sasaki, B. Friman, and  K. Redlich,  Phys. Rev. D {\bf 75}, 054026 (2007);
Phys. Rev. D {\bf 77}, 034024 (2008);
Phys. Rev. Lett. {\bf 99}, 232301 (2007).





\bibitem{fukushima}
K. Fukushima, Phys. Lett. B {\bf 591}, 277 (2004).
\bibitem{PNJL}
C.~Ratti, M.~A.~Thaler and W.~Weise,
Phys.\ Rev.\ D {\bf 73}, 014019 (2006);
C. Sasaki, B. Friman, and  K. Redlich,  Phys. Rev. D {\bf 75}, 074013 (2007);
 K.  Fukushima, J. Phys. G {\bf 36}, 064020 (2009); Phys. Rev. D {\bf 79}, 074015 (2009);
T. Hell,  M. Cristoforetti, and  W. Weise, Phys. Rev. D {\bf 79}, 014022 (2009);
 M. Cristoforetti, T. Hell, B. Klein, and  W. Weise,
e-Print: arXiv:1002.2336 [hep-ph].

\bibitem{Schaefer:2007pw}
  B.-J.~Schaefer, J.~M.~Pawlowski and J.~Wambach,
  Phys.\ Rev.\  D {\bf 76}, 074023 (2007).

\bibitem{Schaefer:2009ui}
  B.-J.~Schaefer, M.~Wagner, J.~Wambach,
  Phys.\ Rev.\  D {\bf 81},   074013 (2010).

\bibitem{Schaefer:2008hk}
  B.-J.~Schaefer, M.~Wagner,
  Phys.\ Rev.\  D {\bf 79},  014018 (2009).




\bibitem{Scavenius:2000qd}
  O.~Scavenius, A.~Mocsy, I.~N.~Mishustin and D.~H.~Rischke,
  Phys.\ Rev.\  C {\bf 64}, 045202 (2001).





\bibitem{qm}
  R.~Friedberg and T.~D.~Lee,
  Phys.\ Rev.\  D {\bf 15}, 1694 (1977);
  Phys.\ Rev.\  D {\bf 16}, 1096 (1977);
  Phys.\ Rev.\  D {\bf 18}, 2623 (1978);
  B.-J.~Schaefer and J.~Wambach,
  Nucl.\ Phys.\  A {\bf 757}, 479 (2005);
  B.-J.~Schaefer, H.-J.~Pirner,
  Nucl.\ Phys.\  {\bf A660 }, 439 (1999).




\bibitem{Schaefer:2006ds}
  B.-J.~Schaefer and J.~Wambach,
  Phys.\ Rev.\  D {\bf 75}, 085015 (2007).

\bibitem{Bowman:2008kc}
  E.~S.~Bowman and J.~I.~Kapusta,
  Phys.\ Rev.\  C {\bf 79}, 015202 (2009).


\bibitem{Gupta:2009fg}
  U.~S.~Gupta and V.~K.~Tiwari,
  arXiv:0911.2464 [hep-ph].

\bibitem{Nickel:2009wj}
  D.~Nickel,
  Phys.\ Rev.\  D {\bf 80}, 074025 (2009).

\bibitem{Kahara:2009sq}
  T.~Kahara and K.~Tuominen,
  Phys.\ Rev.\  D {\bf 80}, 114022 (2009).

\bibitem{Kapusta:2009yd}
  J.~I.~Kapusta and E.~S.~Bowman,
  Nucl.\ Phys.\  A {\bf 830}, 721C (2009).


\bibitem{Nakano:2009ps}
  E.~Nakano, B.-J.~Schaefer, B.~Stokic, B.~Friman and K.~Redlich,
  Phys.\ Lett.\  B {\bf 682}, 401 (2010).

\bibitem{fujii}
H. Fujii and  M. Ohtani,  Phys. Rev. D {\bf 70}, 014016 (2004).

\bibitem{Mizher:2010zb}
  A.~J.~Mizher, M.~N.~Chernodub and E.~S.~Fraga,
  arXiv:1004.2712 [hep-ph].



\bibitem{Pisarski:1983ms} 
R.~D.~Pisarski and F.~Wilczek,
  Phys.\ Rev.\  D {\bf 29}, 338 (1984).

\bibitem{Gasser:1984gg}
  J.~Gasser and H.~Leutwyler,
  Nucl.\ Phys.\  B {\bf 250}, 465 (1985).


\bibitem{Dolan:1973qd}
  L.~Dolan and R.~Jackiw,
  Phys.\ Rev.\  D {\bf 9}, 3320 (1974).

\bibitem{Quiros:1999jp}
  M.~Quiros,
  arXiv:hep-ph/9901312.


\bibitem{Landsman:1986uw}
  N.~P.~Landsman and C.~G.~van Weert,
  Phys.\ Rept.\  {\bf 145}, 141 (1987).

\bibitem{Coleman:1973jx}
S.~Coleman and E.~Weinberg,
Phys.\ Rev.\  D {\bf 7}, 1888 (1973).


\bibitem{Allton:2003vx}
  C.~R.~Allton et al., 
  Phys.\ Rev.\  D {\bf 68}, 014507 (2003).

\bibitem{F1}
F.~Karsch, S.~Ejiri and K.~Redlich, Nucl. Phys. A {\bf 774}, 619 (2006);
S. Ejiri et al., Nucl. Phys. A {\bf 774}, 837 (2006).


\bibitem{Wagner:2009pm}
  M.~Wagner, A.~Walther, B.-J.~Schaefer,
  Comput.\ Phys.\ Commun.\  {\bf 181}, 756 (2010);
  B.-J.~Schaefer, M.~Wagner, J.~Wambach,
  PoS {\bf  CPOD 2009}, 017 (2009), 
  arXiv:0909.0289 [hep-ph].



\bibitem{Ejiri:2005wq}
  S.~Ejiri, F.~Karsch and K.~Redlich,
  Phys.\ Lett.\  B {\bf 633}, 275 (2006).


\bibitem{kurtosis}
  B.~Stokic, B.~Friman and K.~Redlich,
  Phys.\ Lett.\  B {\bf 673}, 192 (2009).




\bibitem{lgt2}
M. Cheng et al., 
Phys. Rev. D {\bf 79}, 074505 (2009). 



\end{thebibliography}
\end{document}